\documentclass[prd,twocolumn,showpacs,preprintnumbers]{revtex4}
\usepackage{hyperref,amssymb,amsmath}
\begin{document}

\title{First--order quasilinear canonical representation of the
characteristic formulation of the Einstein equations}

\author{Roberto G\'omez}
\email{gomez@psc.edu} 
\affiliation{Pittsburgh Supercomputing
Center, 4400 Fifth Ave, Pittsburgh, PA 15213}
\affiliation{Department of Physics and Astronomy, University of
Pittsburgh, Pittsburgh, PA 15260}

\author{Simonetta Frittelli}
\email{simo@mayu.physics.duq.edu}
\affiliation{Department of
Physics, Duquesne University, Pittsburgh, PA 15282}
\affiliation{Department of Physics and Astronomy, University of
Pittsburgh, Pittsburgh, PA 15260}
\begin{abstract}

We prescribe a choice of 18 variables in all that casts the
equations of the fully nonlinear characteristic formulation of
general relativity in first--order quasi-linear canonical form. At
the analytical level, a formulation of this type allows us to make
concrete statements about existence of solutions. In addition, it
offers concrete advantages for numerical applications as it now
becomes possible to incorporate advanced numerical techniques for
first order systems, which had thus far not been applicable to the
characteristic problem of the Einstein equations, as well as in
providing a framework for a unified treatment of the vacuum and
matter problems. This is of relevance to the accurate simulation
of gravitational waves emitted in astrophysical scenarios such as
stellar core collapse.

\end{abstract}

\pacs{04.20.Ex, 04.30.-w, 04.25.Dm}
\preprint{NSF-KITP-03-58}

\maketitle

\section{Introduction}

The characteristic formulation of general relativity due to Bondi and
Sachs~\cite{bondi,sachs}, in particular the approach based on a null
slicing of spacetime with a transverse timelike data surface as proposed
by Tamburino and Winicour~\cite{tamburino}, has been used successfully for
many applications in numerical relativity. It provides a natural framework
for the computation of gravitational radiation signals from isolated
astrophysical sources by purely characteristic evolution~\cite{news},
and as a means of extracting gravitational radiation information from
$3+1$ simulations~\cite{cce}.

It has been used to achieve long-term stable numerical evolutions
of generic single black hole spacetimes~\cite{stable,moving}, it
has also been used to compute the behavior of matter fields around
black holes~\cite{matter,philiptoni,lcone}, and is ideally suited
to the study of gravitational radiation of black hole-neutron star
mergers, the prime candidates for detection by advanced
gravitational wave interferometers~\cite{FinnThorne}. A
modification proposed recently~\cite{particle} incorporates into
the formulation a partial treatment of matter fields with the
specific goal of modeling the capture of a massive object such as
a neutron star by a galactic-size black hole, up to the point
where tidal disruptions become important. Events of this type are
the predominant sources of gravitational radiation which are
expected to fall in the frequency band of LISA.

Additionally, the formulation provides a unique approach to the
post-merger regime of binary black hole coalescence, starting from the
gravitational radiation emitted during a white hole fission~\cite{white},
as illustrated in~\cite{close,close2} in the close-limit of the binary
black hole merger. It has also been used to study the nonlinear generation
of waveforms in single black-hole spacetimes~\cite{philipharm,mode}, where
it has shown the potential to generate a catalog of waveforms, also of
interest for data analysis of space-based gravitational interferometers.

One would expect the stability properties exhibited
by numerical representations of the characteristic
problem~\cite{stable,moving,news,cce,alterprd} to reflect underlying
stability properties at the analytical level, perhaps along the
lines of \cite{friedrich,stewart}. In this regard, in~\cite{sfll} the
linearized equations are cast into a \textit{canonical\/} first--order
form that is suitable for the use of Duff's theorem of existence of
solutions~\cite{duff}. However, the choice of variables of \cite{sfll}
does not allow for an extension of the result to the non-linear case in
any obvious manner. It is desirable to have a quasilinear formulation
of the characteristic equations~\cite{cce,alterprd} in first--order form
that is to the characteristic problem what a first--order formulation is
to a Cauchy problem. Such a formulation could in principle be used as
the starting point to approach relevant issues of stability by means of
energy estimates.

Of particular importance to the numerical implementation of the
characteristic approach is that by writing the system of equations in
first--order quasilinear form, conservative schemes and Godunov-type
shock capturing methods~\cite{leveque,wave} commonly employed to
solve non-linear hyperbolic systems of conservation laws (e.g. the
Euler equations) can be brought to bear on the gravitational field
equations~\cite{fontreview}. High resolution shock capturing methods
have been successfully used in the characteristic formulation for
the evolution of matter fields~\cite{philiptoni,lcone}, but they
have not been applied to the equations for the gravitational fields
themselves, neither in the 3-dimensional case, with the equations in
the form presented in Ref.~\cite{alterprd} or in Ref.~\cite{news},
nor, to the best of our knowledge, to the equations for the
axisymmetric case~\cite{axisymmetric}. The approach developed by Pons
et.~al~\cite{pons} extends the use of special relativistic Riemann
solvers, developed for special relativistic flows~\cite{marti}, through
local coordinate transformations, thereby making them applicable to
general relativistic problems as well. An obvious impediment to a unified
treatment, as proposed in Pons et. al.~\cite{pons}, has been the lack
of a first order representation of the characteristic formulation.

This leads to a subtle point deserving of clarification: true
shocks do not form in the gravitational field, thus strictly
speaking, shock capturing methods are necessary for the matter
evolution equations only. In the presence of matter however, given
that the matter fields act as sources for the metric fields,
shocks and steep gradients in the matter sources lead invariably
to the appearance of short-scale spatial features in the
gravitational fields. The centered finite difference schemes used
in the characteristic
codes~\cite{axisymmetric,news,matter,alterprd} are not capable of
resolving these short scales, and invariably generate high
frequency noise. In a perverse twist, the more successful a shock
capturing method is at resolving short-scale matter features, the
more it compounds the problem of integrating the metric fields
across these features by means of standard centered difference
schemes. The effect is a classical illustration of the error one
finds in advecting a pulse using a non-shock capturing centered
difference method, e.g. Lax-Wendroff, where spurious oscillations
appear on the trailing edge of the numerical solution. With a
convergent scheme, these unwanted oscillations decrease with
increasing resolution and vanish in the continuum limit, but are
present in all simulations with practical grid sizes. Attempts to
filter out the noise lead to a spreading of the pulse. In the
problem at hand, it can lead to an unacceptable trade-off between
the accuracy of the matter fields evolution vs. that of the metric
fields. This effect is mentioned in recent work by Siebel
et.al.~\cite{siebel}, who perform characteristic evolution in
axisymmetric stellar core collapse. In their work, the matter
fields are solved via shock capturing methods~\cite{philiptoni}
while the metric evolution is treated by centered, second order
accurate finite difference schemes~\cite{axisymmetric}, with
modifications along the lines of~\cite{alterprd}. This is a
difficult numerical task: in the core collapse scenario, a strong
shock wave forms after bounce, where all matter fields are
discontinuous. The dynamics of the collapse are correctly solved
and these discontinuities are accurately tracked~\cite{siebel}.
However, the discontinuities in the matter field lead to
unavoidable discontinuities in the first derivatives of the metric
fields. In the traditional scheme, second derivatives of the
metric are need to compute the gravitational radiation, and this
is where numerical noise is generated, the main difficulty pointed
out in~\cite{siebel}. A form of the equations without second order
derivatives would provide a solution to this problem, and thus be
particularly relevant to astrophysical problems with shocks. In
the absence of such a formulation, filtering of the unwanted noise
on a fixed grid has been shown not to provide sufficient accuracy,
and the conclusion drawn in~\cite{siebel} is that unless the short
scale gravitational field features are adequately resolved and the
resulting spurious oscillations eliminated, it is difficult to
extract more accurate gravitational signals, only the main
features of the signal being obtained to good accuracy.

It is generally accepted that it will be necessary to incorporate adaptive
mesh refinement techniques in order to achieve well-resolved simulations
with matter sources~\cite{particle,siebel}. Preliminary steps in these
direction have been taken, with the implementation of an adaptively
refined code for the model problem of Einstein-Klein-Gordon fields in
spherical symmetry~\cite{pretorius}. This task can also be more easily
addressed when both the gravitational field equations and the matter
equations are written in first order differential form. We should also
point out another difference between the matter and gravitational field
equations. Since the matter evolution equations express the conservation
of the stress-energy tensor, they can be easily expressed in conservative
form~\cite{philiptoni}. The gravitational field evolution equations,
on the other hand, need not be put in conservative form, as they can
be treated with volume preserving methods which apply equally to first
order quasilinear systems and are also available~\cite{curvewave}.

Here we introduce a set of variables and auxiliary equations that
casts the nonlinear gravitational field equations into a
quasilinear, first--order representation of the Einstein equations
in the null cone formalism that takes Duff's canonical form and
provides a bridge to potential adaptations of Cauchy methods to
the characteristic problem for the Einstein equations. The
Tamburino-Winicour version of the Bondi-Sachs characteristic
problem reduced to first--order in angular derivatives is reviewed
in Section~\ref{sec:2} and is taken as the starting point for the
remainder of the work. In Section~\ref{sec:focf} the first--order
quasilinear canonical characteristic form of the equations is
derived. We offer concluding remarks in Section~\ref{sec:4}.

\section{The characteristic problem of the Einstein equations}\label{sec:2}

As in Refs.~\cite{news,cce,alterprd}, we use coordinates based
upon a family of outgoing null hypersurfaces. We let $u$ label
these hypersurfaces, $x^A$ $(A=2,3)$ label the null rays and $r$
be a surface area coordinate, such that in the
$x^\alpha=(u,r,x^A)$ coordinates the metric takes the Bondi-Sachs
form~\cite{bondi,sachs}
\begin{eqnarray}
   ds^2 & = & -\left( e^{2\beta} \frac{V}{r} - r^2h_{AB}U^AU^B \right)du^2
              - 2e^{2\beta}dudr  \nonumber \\
              &-& 2r^2 h_{AB}U^Bdudx^A + r^2h_{AB}dx^Adx^B,
\label{eq:bmet}
\end{eqnarray}
\noindent where $h_{AB}$ is conformal to the metric of the
sections of fixed value of $r$ on the null slice, and
$\det(h_{AB})=\det(q_{AB})$, with $q_{AB}$ a unit sphere metric.
We define the inverse by $h^{AB}h_{BC}=\delta^A_C$. By
representing tensors in terms of spin-weighted
variables~\cite{news}, the conformal metric $h_{AB}$ is completely
encoded in the complex function $J\equiv\frac{1}{2}h_{AB}q^Aq^B$,
where $q^A$ is a complex dyad such that $q^A\bar{q}_A = -2$ and
$q^Aq_A=0$ (an overbar denotes complex conjugation).
The remaining dyad component of the conformal metric,
given by the real function $K=\frac{1}{2}h_{AB}q^A\bar q^B$, is
determined by $K^2=1+J\bar J$ as a consequence of the determinant
condition.
Additionally, we define $U\equiv U_Aq^A$. Angular derivatives of
tensor components are in turn expressed in terms of $\eth$ and
$\bar\eth$ operators~\cite{eth}.

The equations for the characteristic (or null cone) formulation follow
from projections of the Ricci tensor normal and tangent to the null
slices~\cite{sachs}. The resulting \textit{main equations} arrange into
a hierarchy, splitting into a set of \textit{hypersurface equations,\/}
which involve only derivatives on the null cone, and \textit{evolution
equations} involving derivatives with respect to the retarded time
$u$. In particular, $R_{rr}=0$ provides an equation for $\beta_{,r}$
in terms of $J$, while $R_{rA}q^A=0$ gives $U,_{rr}$ in terms of $J$ and
$\beta$, and the trace $R_{AB}h^{AB}=0$ yields $V_{,r}$ in terms of $J$,
$\beta$ and $U$. Finally, $R_{AB}q^Aq^B$ supplies the evolution equation
for $J$.  The remaining four components of the Ricci tensor vanish as a
consequence of these six in the following sense. By virtue of the Bianchi
identities, the component $R_{ur}=0$ is trivially satisfied wherever the
\textit{main equations} are satisfied, whereas the remaining components
$R_{uu}=0$ and $R_{uA}q^A=0$ are propagated radially on the null slices
if they hold on a surface $r=r_0$. Thus $R_{uu}=0$ and $R_{uA}q^A=0$
(the \textit{supplementary conditions}) can be viewed as constraints on
the data at $r=r_0$. In the following, we ignore these constraints.
We will also ignore matter source terms, although including them is
straightforward (see~\cite{matter}), in the interest of keeping the
presentation concise.

For the present derivation, we find it convenient to start from a
relatively recent representation of the characteristic
formulation~\cite{alterprd} which casts the system into
first--order form in the angular derivatives, and mixed
first--second--order form in the radial derivatives --as opposed to
the standard, mixed--order form~\cite{cce,news}. We depart however
from~\cite{alterprd} in our choice of fundamental variables in
order to facilitate the presentation in Sec.~\ref{sec:focf}. In
this {\it partially reduced\/} form, the complete system of main
equations of the characteristic formulation consists of a complex
evolution equation for the conformal metric
\begin{equation}
     2 \left(rJ\right)_{,ur}
    - \left((1+r\tilde{W})\left(rJ\right)_{,r}\right)_{,r}
     = {\cal D} + J_{H} + J P_u ,
    \label{eq:J}
\end{equation}
\noindent namely Eq.~(16) of Ref.~\cite{alterprd}, and the
hierarchy of hypersurface equations and auxiliary definitions as
follows:
\begin{eqnarray}
   \nu_{,r} &=& \bar\eth J_{,r}
   \label{eq:nu} \\
   \mu_{,r} &=& \eth J_{,r}
   \label{eq:mu}
\\
\beta_{,r} &=& \frac{r}{8} \left( J_{,r} \bar J_{,r} -
K^2_{,r}\right)
   \label{eq:beta}
\\
B_{,r} &=& \eth\beta_{,r}
   \label{eq:B}
\end{eqnarray}
\begin{eqnarray}
     (r^2 Q)_{,r}  &=& 2 r^2 B_{,r} - 4r B
   +  r^2 \bigg[ - K ( k_{,r} + \nu_{,r} ) \nonumber \\ &&
                 + \bar \nu J_{,r} + \bar J \mu_{,r} + \nu K_{,r}
                 + J \bar k_{,r} - J_{,r} \bar k
          \bigg] \nonumber \\
     && + \frac{r^2}{2K^2} \left[ \bar \nu (     J_{,r} -      J^2 \bar J_{,r})
                                 +     \mu (\bar J_{,r} - \bar J^2      J_{,r})
                           \right]
\label{eq:Q}
\\
r^2 U_{,r}  &=& e^{2\beta} \left( K Q - J \bar Q \right)
     \label{eq:U}
     \\
      (r^2 \tilde{W})_{,r}  &=&  \frac{1}{2} e^{2\beta} {\cal R} - 1
    + r \bar \eth U + r \eth \bar U
    + \frac{r^2}{4} ( \bar\eth U_{,r} + \eth \bar U_{,r} )
  \nonumber \\
   &+& e^{2\beta} \Big( -K \left( \bar\eth B + B \bar B\right) \nonumber\\
   &+& \frac{1}{2} [  \bar J \left(      \eth      B +      B^2 \right)
                     +     J \left( \bar \eth \bar B + \bar B^2 \right) ]
\nonumber \\
    && + \frac{1}{2} [  (     \nu -      k) \bar B
                      + (\bar \nu - \bar k)      B ]
\Big)
    \nonumber \\
    &-& e^{-2 \beta} \frac{r^4}{8} \Big[
     \bar U_{,r} \left( K      U_{,r} +      J \bar U_{,r} \right)
  \nonumber \\
    && +\
          U_{,r} \left( K \bar U_{,r} + \bar J      U_{,r} \right)
          \Big]
 \label{eq:W}
\end{eqnarray}
\noindent These are Eqs.(21)-(26) of~\cite{alterprd} with the
identification $\mu\equiv \eth J$. In Eq.~(\ref{eq:J}), the left--hand
side is a characteristic representation of a wave operator of second
differential order in $r$ and $u$, and we have split the right hand-side
into three terms, according to usage. The symbol $P_u$
singles out the only term where a retarded time derivative appears in
any right--hand side, namely $J,_u$:
\begin{equation}
  P_{u} \equiv \frac{r}{K}
  \left[     J_{,u} \left( \bar J_{,r} K - \bar J K_{,r}\right)
+       \bar J_{,u} \left(      J_{,r} K -      J K_{,r}\right) \right]
 \label{eq:P_u}
\end{equation}
\noindent The occurrence of this retarded time derivative in first
order will be found to be critical to the developments that follow in
Section~\ref{sec:focf}.  The symbol $J_{H}$ stands for the following
collection of terms:
\begin{eqnarray}
J_{H}&=& \frac{e^{2 \beta}}{r}
  \bigg (
          - K ( \mu \bar B + 2 k B - \nu B)
          + B (\bar J \mu + J \bar \nu )
\nonumber \\
    & & +  J ( \bar B k - B  \bar k )
        + J \big[
                   - 2 K ( \eth \bar B + B \bar B )
\nonumber \\
    & &           +      J ( \bar \eth \bar B + \bar B^2 )
                  + \bar J (      \eth B             + B^2 )
            \big]
  \bigg)
\nonumber \\
      & & + \frac{r^3}{2} e^{-2 \beta} \bigg(
\left( K U_{,r} + J \bar U_{,r} \right)^2
\nonumber \\
    & & - \frac{J}{2}
\left[ ( K U_{,r} + J \bar U_{,r} ) \bar U_{,r}
+ (K \bar U_{,r} + \bar J U_{,r} ) U_{,r}
\right] \bigg)
\nonumber \\
      & & - \frac{\nu}{2} ( r      U_{,r} + 2      U)
          - \frac{\mu}{2} ( r \bar U_{,r} + 2 \bar U)
\nonumber \\
    & & + \frac{J}{2} ( r \bar \eth U_{,r} + 2 \bar \eth U)
        - \frac{J}{2} ( r  \eth \bar U_{,r} + 2 \eth \bar U)
\nonumber \\
      & & + (1-K) ( r  \eth U_{,r} + 2 \eth U)
           - \frac{r}{2} J_{,r} ( \eth \bar U + \bar \eth U)
\nonumber \\
      & & + \frac{r}{2} ( \bar U \mu + U \nu )
       (J \bar J_{,r} - \bar J J_{,r} )
     - r \bar U \mu_{,r}
     - r U \nu_{,r}
\nonumber \\
&+& r ( J_{,r} K -  J K_{,r} )
      \bigg ( k \bar U  + \bar k U
        + K ( \bar \eth U - \eth \bar U )
\nonumber \\
& &
        +  J \bar \eth \bar U - \bar J \eth U
      \bigg)
        - 8 J (1 + r \tilde{W}) \beta_{,r}
\label{eq:J_H}
\end{eqnarray}
\noindent Any remaining terms are collected into the symbol $\cal
D$ for notational convenience:
\begin{equation}
{\cal D} = -K ( \eth r U_{,r} + 2 \eth U)
+ \frac{2 e^{2\beta} }{r} (\eth B + B^2) - (r\tilde{W})_{,r} J .
\end{equation}
\noindent We also have
\begin{equation}
  {\cal R} = 2 K + \frac{1}{2} \big[
          \bar \eth \left(     \nu -      k \right)
        +      \eth \left(\bar \nu - \bar k \right)
  \big]
  + \frac{1}{4K} \left( |\mu|^2 - |\nu|^2 \right) ,
   \label{eq:Ricci}
\end{equation}
\noindent and use the symbol $k$ throughout for
\begin{equation}\label{k}
  k\equiv \frac{\mu\bar{J}+J\bar{\nu}}{2K}.
\end{equation}
\noindent The symbol $\tilde{W}$ is simply a renaming of the original
metric function $V$ that is regular at $r=\infty$, being defined by
$V\equiv r+r^2\tilde{W}$. The symbol $Q$ is a first--order variable
encoding the radial derivative of $U$, and is \textit{defined} by
Eq.~(\ref{eq:U}), which acts as a hypersurface equation for $U$. In
addition, $\nu, \mu$ and $B$ are first--order variables of spin weights
1, 3 and 1, respectively, used to reduce the differential order of the
angular derivatives appearing in the original characteristic equations,
and are defined by
\begin{equation}\label{munuB}
\nu\equiv\bar\eth J, \qquad \mu\equiv\eth J,\qquad
B\equiv\eth\beta.
\end{equation}
\noindent With the definitions~(\ref{k})-(\ref{munuB}),
Eq.~(\ref{eq:J_H}) follows from Eq.~(25) of~\cite{alterprd}.
Equations~(\ref{eq:J})-(\ref{eq:W}) as they stand contain no second--order
derivatives in the retarded time or the angular coordinates. Additionally,
all appearances of $U,_r$ and $\tilde{W},_r$ in the right--hand sides
represent angular derivatives and undifferentiated terms by virtue
of~(\ref{eq:U}) and (\ref{eq:W}). The system is still of second
differential order overall, because of the presence of second--order
derivatives of $J$, but its value resides in the fact that it exhibits
remarkable numerical stability properties~\cite{alterprd}, raising the
question of whether a full reduction to proper first order may further
enhance the numerical stability.

\section{Reduction of the Tamburino-Winicour system to first--order
quasilinear form} \label{sec:focf}

Our immediate goal is to write the full nonlinear equations in a
quasi-linear first--order form in the strict sense, that is:
\begin{equation}
  A^\alpha(u,r,x^A,v) v_{,\alpha}+s(u,r,x^A,v) =0,
\end{equation}
\noindent where $v$ represents the set of all dependent variables,
the index $\alpha$ runs over all spacetime coordinates and where the
matrices $A^\alpha$ and the vector of source terms $s$ depend on the
coordinates and the undifferentiated variables $v$. Since all remaining
second--order terms contain $J,_r$, and because \textit{the only nonlinear
terms in first-derivatives are quadratic and contain $J,_r$ as a factor}
-- see Eq.~(\ref{eq:P_u})--, this can be accomplished if an appropriate
$r-$derivative of the fundamental field $J$ is re-defined as an additional
fundamental variable, and there is any number of different acceptable
re-definitions, one of which was used in~\cite{sfll}. Proceeding along
the lines of Ref.~\cite{alterprd}, we define
\begin{equation}
    H \equiv (r J)_{,r},
    \label{eq:auxdef}
\end{equation}
\noindent which has spin weight $2$. No other radial derivatives are
necessary to convert Eqs.~(\ref{eq:J})-(\ref{eq:W}) down to first--order
\textit{quasilinear\/} form.  With this definition, the left--hand side
of Eq.~(\ref{eq:J}) becomes $2H,_u - \left( ( 1+ r \tilde{W}) H \right),_r$.
In the process,
however, the term $J,_u$ in the right--hand side of Eq.~(\ref{eq:J})
is promoted to the principal symbol of the system, with the consequence
that the evolution equation involves the retarded time derivatives of two
complex variables ($H$ and $J$), instead of just one. It is unclear at
this point whether Eq.~(\ref{eq:J}) would determine the evolution of $H$
or of $J$. (The difficulty does not arise if one linearizes the equations
before reducing to first--order form, as was done in~\cite{sfll}.) In
fact, we know of no solution--generating process for the system at this
point. So we proceed as follows.

In order to avoid the occurrence of the retarded-time derivative of
$J$ in the right--hand side of Eq.~(\ref{eq:J}) we \textit{define it\/}
as an additional fundamental variable:
\begin{equation}
F \equiv  J_{,u}, \label{eq:Fdef}
\end{equation}
\noindent which has spin weight $2$. This is at first counter-intuitive:
the equations are already first order in $\partial_u$, so defining
the $u-$derivative as a new variable might not appear necessary,
or even consistent. However, in the following we show that by
defining this additional variable, the characteristic equations
take the canonical hierarchical form needed for the existence of a
solution from characteristic data~\cite{duff}. With the definitions
(\ref{eq:auxdef})-(\ref{eq:Fdef}), the original evolution equation,
Eq.~(\ref{eq:J}), is interpreted as a wave equation for $H$,
shown below as Eq.~(\ref{eq:fqH}). From (\ref{eq:Fdef}) we have
$(rF)_{,r}=(rJ),_{ur}$, which yields a hypersurface equation for $F$,
namely Eq.~(\ref{eq:fqF}) below. With this we can finally write the
system in the form
\begin{subequations}\label{finalsystem}
\begin{eqnarray}
  J_{,r} &=& \frac{1}{r} (H - J) ,
  \label{eq:fqJ}
  \\
  \mu_{,r} &=& \frac{1}{r} (\eth H - \mu ) ,
  \label{eq:fqmu}
  \\
  \nu_{,r} &=& \frac{1}{r} (\bar\eth H - \nu ) ,
  \label{eq:fqnu}
  \\
\beta_{,r} &=& \frac{r}{8} \left( J_{,r} \bar J_{,r} -
K^2_{,r}\right)
  \\
  8 r B_{,r} &=& r \mu_{,r} ( \bar H - \bar J ) + r \bar\nu_{,r} ( H - J)
             \nonumber \\
  &-& \frac{1}{K} \left[ \bar J ( H - J ) + J ( \bar H - \bar J ) \right]
                  r k_{,r}
\label{eq:fqB}
\end{eqnarray}
\begin{eqnarray}
     (r^2 Q)_{,r}  &=& 2 r^2 B_{,r} - 4r B
   +  r^2 \bigg[ - K ( k_{,r} + \nu_{,r} ) \nonumber \\ &&
                 + \bar \nu J_{,r} + \bar J \mu_{,r} + \nu K_{,r}
                 + J \bar k_{,r} - J_{,r} \bar k
          \bigg] \nonumber \\
     &+& \frac{r^2}{2K^2} \left[ \bar \nu (     J_{,r} -      J^2 \bar J_{,r})
                                 +     \mu (\bar J_{,r} - \bar J^2      J_{,r})
                           \right]
\\
   r^2 U_{,r}  &=& e^{2\beta} \left( K Q - J \bar Q \right)
  \label{eq:fqU} \\
     (r^2 \tilde{W})_{,r}  &=& \frac{1}{2} e^{2\beta} {\cal R} - 1
     + r \bar \eth U + r \eth \bar U
          \nonumber \\
   & + & \frac{1}{4} \left( \bar \eth \left[ e^{2\beta} ( K Q - J \bar Q) \right]
                               + \eth \left[ e^{2\beta} ( K \bar Q - \bar J Q) \right]
                     \right)
   \nonumber \\
   &+& e^{2\beta} \Big( -K \left( \bar\eth B + B \bar B\right) \nonumber\\
   &+& \frac{1}{2} [  \bar J \left(      \eth      B +      B^2 \right)
                     +     J \left( \bar \eth \bar B + \bar B^2 \right) ]
\nonumber \\
    && + \frac{1}{2} [  (     \nu -      k) \bar B
                      + (\bar \nu - \bar k)      B ]
\Big)
              \nonumber\\
   &-& \frac{e^{2\beta}}{8}
           \left[       Q (K \bar Q - \bar J      Q)
                 + \bar Q (K      Q -      J \bar Q) \right]
  \label{eq:fqW}
\\
2(r F)_{,r} &=& \left( (1 + r\tilde{W}) H \right)_{,r} + {\cal D} + J_H + J P_u,
\label{eq:fqF}
\end{eqnarray}
for the hypersurface equations and
\begin{equation}
  2H_{,u} - [( 1 + r\tilde{W} ) H]_{,r}
  =  {\cal D} + J_H + J P_u ,
\label{eq:fqH}
\end{equation}
\end{subequations}
for the evolution equation, where
\begin{eqnarray}
   P_{u} &=& F(\bar H - \bar J) + \bar F(H - J)
   \nonumber \\
   &-&\left(\frac{F \bar J + \bar F J}{2K^2}\right)
   \left[(H-J)\bar{J} + J (\bar{H}-\bar{J})\right]
\end{eqnarray}
\noindent Eqs.~(\ref{eq:fqmu}), (\ref{eq:fqnu}), (\ref{eq:fqB})
and (\ref{eq:fqF}) arise from taking an $r-$derivative of
(\ref{munuB}) and commuting the derivatives in the right--hand
side, as usual. All the radial derivatives indicated in the
right--hand side of Eqs.~(\ref{eq:fqB})-(\ref{eq:fqH}) can be
substituted, in turn, by quantities computed previously in the
hierarchy, i.e. the right--hand sides of the equations can be
expressed purely in terms of the undifferentiated fundamental
variables and their angular derivatives. The substitutions have
been left indicated for the sake of brevity, noting only that the
derivatives of $k$ (which is not part of the hierarchy) are given by
\begin{eqnarray}
  k_{,r} &=& \frac{1}{2 K}
             \left(   \bar J      \mu_{,r}
                    +      J \bar \nu_{,r}
                    + \bar \nu      J_{,r}
                    +      \mu \bar J_{,r}
             \right)
          - \frac{k K_{,r}}{K}
\nonumber \\
  \bar \eth k &=&
  \frac{1}{2K}
   \left( \bar J \bar \eth \mu + J \bar \eth \bar \nu + \mu^2 + \nu \bar \nu
   \right)
  - \frac{k \bar k}{K}
  \label{eq:fqk}
\end{eqnarray}
Equations~(\ref{eq:fqJ})-(\ref{eq:fqF}) can be viewed as propagation
equations along the radially outgoing null geodesics, with
Eq.~(\ref{eq:fqH}) advancing the radial derivative of the spherical
metric function $J$ in time.

In this first--order formulation of the null cone approach,
Eqs.~(\ref{eq:fqJ})-(\ref{eq:fqH}), the boundary data at $r=r_0$ consists
of the values of $J$, $\beta$, $Q$, $U$ and $\tilde W$, with the values
of $\mu$, $\nu$, $B$ and $F$ following from the boundary value of $J$
as per Eqs.~(\ref{munuB}) and (\ref{eq:auxdef}). The consistency
conditions, imposed at $r=r_0$, are propagated to the interior by
Eqs.~(\ref{eq:fqmu}), (\ref{eq:fqnu}), (\ref{eq:fqB}) and (\ref{eq:fqF}).
The initial data for the system  (\ref{eq:fqJ})-(\ref{eq:fqH}) at $u=u_0$
are the values of $H(r,x^A)$, representing  the shear of the conformal
metric of the spheres, given on the entire initial hypersurface.
The conformal metric function $J$ on the initial hypersurface
follows by integration of $H$ as per Eq.~(\ref{eq:fqJ}), with the
integration constant provided by the value of $J$ at the boundary.
Eqs.~(\ref{eq:fqmu}) through (\ref{eq:fqF}) in turn provide initial values
for $\mu$, $\nu$, $k$, $\beta$, $B$, $Q$, $U$, $\tilde W$ and $F$, while
Eq.~(\ref{eq:fqH}) propagates $H$ forward in retarded time. At this point,
the process can be repeated, and the entire space-time exterior to the
time-like data surface can be computed.

This solution--generating process lies at the basis of Duff's theorem of
existence and uniqueness of solutions to characteristic problems --that
is: problems for hyperbolic systems of equations where data is prescribed
on a characteristic surface. From the analytical point of view, as it
stands, the system of Eqs.~(\ref{finalsystem}) has the form
\begin{eqnarray}
\partial_u q + N\partial_r q
&=& L^1(\eth q,\bar\eth q, \eth w,\bar\eth w, q,w)\\
\partial_r w + M\partial_r q
&=& L^2(\eth q,\bar\eth q, \eth w,\bar\eth w, q,w)
\end{eqnarray}
\noindent where $q\equiv H$, $w \equiv
(J,\mu,\nu,\beta,B,U,Q,\tilde{W},F)$, and $N$ and $M$ are certain matrices
of dimension $2\times 2$ and $14\times 2$ respectively, depending
on the undifferentiated variables. A trivial change of variable $F\to
F-(1+r\tilde{W})H/(2r)$ puts the system of
equations~(\ref{eq:fqJ})-(\ref{eq:fqH})
into a 18-dimensional first--order canonical quasi-linear form as
defined by Duff~\cite{duff}, for 18 variables of which two ($H$) are
\textit{normal} and the remaining 16 ($J,\mu,\nu,\beta,B,U,Q,\tilde{W},F$)
are \textit{null}, and with four complex constraints ${\cal C}_i$ on the
surface $r=r_0$ which are trivially preserved by the solution--generating
process in the form $\partial_r{\cal C}_i = 0$. This means that the
system (\ref{finalsystem}) satisfies the conditions for Duff's theorem,
and therefore existence and uniqueness follow from null and normal data
in a manner analogous to Cauchy problems. This is not a trivial result,
as readers should note that the same cannot be said if $J,_u$ is not
defined as a fundamental variable.

\section{Conclusion}\label{sec:4}

The system of equations~(\ref{eq:fqJ})-(\ref{eq:fqH}) casts the
Tamburino-Winicour version of the Bondi-Sachs characteristic initial
value problem into a standard first--order quasilinear form.

A novel feature of this formulation is the introduction of a $u-$
derivative of the 2-sphere metric, $h_{AB,u}$, as a fundamental
variable. The necessity of this step arises only in the full non-linear
characteristic problem, signaling the fact that the linearization and
``canonization'' operations (i.e. the reduction to a hierarchy as per the
Bondi Sachs construction) do not commute in the case of the characteristic
problem of the Einstein equations.  This form of the equations opens
the possibility of further studies at the analytic level, specifically
to look for the existence of estimates of the solution in terms of the
data on the initial characteristic surface and the data on the surface
of fixed radius $r_0$.

In a much broader context, we have constructively shown here that only 18
variables are needed in order to achieve a first--order formulation of the
Einstein equations suitable for numerical generation of solutions. And,
most remarkably, the resulting nonlinearities are of the quasilinear
type. By contrast, with regards to the Cauchy problem of the Einstein
equations, Alekseenko and Arnold~\cite{alekseenko} show that as many as
eight variables are actually needed in addition to the six three-metric
components and the six extrinsic curvature components in order to obtain
a full first--order reduction of the ADM equations~\cite{yorksources}.
This yields a total of at least 20 variables for the first--order
Cauchy problem, with the drawback that the resulting non-linearities
are genuine (not of quasilinear type). In order to remove the genuine
non-linearities, all 18 space-derivatives of the three-metric must
be added as fundamental variables, with the result that the generic
quasilinear first--order reduction of the 3+1 Einstein equations requires
a number of 30 variables. From this perspective, the fact that only 18
variables in all are actually sufficient for a first--order quasilinear
solution--generating process for the Einstein equations from given data
is both surprising and intriguing.

But perhaps the most important feature of the first--order
formulation Eqs. (\ref{eq:fqJ})--(\ref{eq:fqH}) is its potential
for accurately handling discontinuities in the first derivatives
of the metric. This is very relevant to simulations of systems of
astrophysical interest involving shock waves, such as stellar core
collapse, where discontinuities in the matter fields arise and are
transmitted to the first derivatives of the metric. In such
systems, an accurate treatment of those discontinuities is
essential to ensure the quality of the waveforms obtained. The
quasilinear form of the equations is thus ideal for a unified
treatment of the gravitational and matter evolution equations,
with the introduction of more advanced numerical algorithms, in
particular along the lines of~\cite{philiptoni,lcone,pons,siebel}.
Work in these directions is currently in progress. Results of the
application of the system of equations introduced here to the
numerical characteristic effort will be reported elsewhere.

\begin{acknowledgments}

RG thanks the Kavli Institute for Theoretical Physics for
hospitality. This research was supported by the National Science
Foundation under grants No. PHY-0070624 to Duquesne University,
No. PHY-0135390 to Carnegie Mellon University, and No. PHY-9907949 to
the University of California at Santa Barbara.

\end{acknowledgments}


\begin{thebibliography}{34}
\expandafter\ifx\csname natexlab\endcsname\relax\def\natexlab#1{#1}\fi
\expandafter\ifx\csname bibnamefont\endcsname\relax
  \def\bibnamefont#1{#1}\fi
\expandafter\ifx\csname bibfnamefont\endcsname\relax
  \def\bibfnamefont#1{#1}\fi
\expandafter\ifx\csname citenamefont\endcsname\relax
  \def\citenamefont#1{#1}\fi
\expandafter\ifx\csname url\endcsname\relax
  \def\url#1{\texttt{#1}}\fi
\expandafter\ifx\csname urlprefix\endcsname\relax\def\urlprefix{URL }\fi
\providecommand{\bibinfo}[2]{#2}
\providecommand{\eprint}[2][]{\url{#2}}

\bibitem[{\citenamefont{Bondi et~al.}(1962)\citenamefont{Bondi, van~der Burg,
  and Metzner}}]{bondi}
\bibinfo{author}{\bibfnamefont{H.}~\bibnamefont{Bondi}},
  \bibinfo{author}{\bibfnamefont{M.~J.~G.} \bibnamefont{van~der Burg}},
  \bibnamefont{and} \bibinfo{author}{\bibfnamefont{A.~W.~K.}
  \bibnamefont{Metzner}}, \bibinfo{journal}{Proc. R. Soc. London}
  \textbf{\bibinfo{volume}{A269}}, \bibinfo{pages}{21} (\bibinfo{year}{1962}).

\bibitem[{\citenamefont{Sachs}(1962)}]{sachs}
\bibinfo{author}{\bibfnamefont{R.~K.} \bibnamefont{Sachs}},
  \bibinfo{journal}{Proc. R. Soc. London} \textbf{\bibinfo{volume}{A270}},
  \bibinfo{pages}{103} (\bibinfo{year}{1962}).

\bibitem[{\citenamefont{Tamburino and Winicour}(1966)}]{tamburino}
\bibinfo{author}{\bibfnamefont{L.~A.} \bibnamefont{Tamburino}}
  \bibnamefont{and} \bibinfo{author}{\bibfnamefont{J.}~\bibnamefont{Winicour}},
  \bibinfo{journal}{Phys. Rev.} \textbf{\bibinfo{volume}{150}},
  \bibinfo{pages}{1039} (\bibinfo{year}{1966}).

\bibitem[{\citenamefont{Bishop et~al.}(1997)\citenamefont{Bishop, G\'omez,
  Lehner, Maharaj, and Winicour}}]{news}
\bibinfo{author}{\bibfnamefont{N.~T.} \bibnamefont{Bishop}},
  \bibinfo{author}{\bibfnamefont{R.}~\bibnamefont{G\'omez}},
  \bibinfo{author}{\bibfnamefont{L.}~\bibnamefont{Lehner}},
  \bibinfo{author}{\bibfnamefont{M.}~\bibnamefont{Maharaj}}, \bibnamefont{and}
  \bibinfo{author}{\bibfnamefont{J.}~\bibnamefont{Winicour}},
  \bibinfo{journal}{Phys. Rev. D} \textbf{\bibinfo{volume}{56}},
  \bibinfo{pages}{6298} (\bibinfo{year}{1997}).

\bibitem[{\citenamefont{Bishop et~al.}(1996)\citenamefont{Bishop, G\'omez,
  Lehner, and Winicour}}]{cce}
\bibinfo{author}{\bibfnamefont{N.~T.} \bibnamefont{Bishop}},
  \bibinfo{author}{\bibfnamefont{R.}~\bibnamefont{G\'omez}},
  \bibinfo{author}{\bibfnamefont{L.}~\bibnamefont{Lehner}}, \bibnamefont{and}
  \bibinfo{author}{\bibfnamefont{J.}~\bibnamefont{Winicour}},
  \bibinfo{journal}{Phys. Rev. D} \textbf{\bibinfo{volume}{54}},
  \bibinfo{pages}{6153} (\bibinfo{year}{1996}).

\bibitem[{\citenamefont{G\'omez
  et~al.}(1998{\natexlab{a}})\citenamefont{G\'omez, Lehner, Marsa, Winicour
  et~al.}}]{stable}
\bibinfo{author}{\bibfnamefont{R.}~\bibnamefont{G\'omez}},
  \bibinfo{author}{\bibfnamefont{L.}~\bibnamefont{Lehner}},
  \bibinfo{author}{\bibfnamefont{R.}~\bibnamefont{Marsa}},
  \bibinfo{author}{\bibfnamefont{J.}~\bibnamefont{Winicour}},
  \bibnamefont{et~al.}, \bibinfo{journal}{Phys. Rev. Lett.}
  \textbf{\bibinfo{volume}{80}}, \bibinfo{pages}{3915}
  (\bibinfo{year}{1998}{\natexlab{a}}).

\bibitem[{\citenamefont{G\'omez
  et~al.}(1998{\natexlab{b}})\citenamefont{G\'omez, Lehner, Marsa, and
  Winicour}}]{moving}
\bibinfo{author}{\bibfnamefont{R.}~\bibnamefont{G\'omez}},
  \bibinfo{author}{\bibfnamefont{L.}~\bibnamefont{Lehner}},
  \bibinfo{author}{\bibfnamefont{R.~L.} \bibnamefont{Marsa}}, \bibnamefont{and}
  \bibinfo{author}{\bibfnamefont{J.}~\bibnamefont{Winicour}},
  \bibinfo{journal}{Phys. Rev. D} \textbf{\bibinfo{volume}{57}},
  \bibinfo{pages}{4778} (\bibinfo{year}{1998}{\natexlab{b}}).

\bibitem[{\citenamefont{Bishop et~al.}(1999)\citenamefont{Bishop, G\'omez,
  Lehner, Maharaj, and Winicour}}]{matter}
\bibinfo{author}{\bibfnamefont{N.~T.} \bibnamefont{Bishop}},
  \bibinfo{author}{\bibfnamefont{R.}~\bibnamefont{G\'omez}},
  \bibinfo{author}{\bibfnamefont{L.}~\bibnamefont{Lehner}},
  \bibinfo{author}{\bibfnamefont{M.}~\bibnamefont{Maharaj}}, \bibnamefont{and}
  \bibinfo{author}{\bibfnamefont{J.}~\bibnamefont{Winicour}},
  \bibinfo{journal}{Phys. Rev. D} \textbf{\bibinfo{volume}{60}},
  \bibinfo{pages}{024005} (\bibinfo{year}{1999}).

\bibitem[{\citenamefont{Papadopoulos and Font}(2000)}]{philiptoni}
\bibinfo{author}{\bibfnamefont{P.}~\bibnamefont{Papadopoulos}}
  \bibnamefont{and} \bibinfo{author}{\bibfnamefont{J.~A.} \bibnamefont{Font}},
  \bibinfo{journal}{Phys. Rev. D} \textbf{\bibinfo{volume}{61}},
  \bibinfo{pages}{024015} (\bibinfo{year}{2000}).

\bibitem[{\citenamefont{Siebel et~al.}(2002)\citenamefont{Siebel, Font, Muller,
  and Papadopoulos}}]{lcone}
\bibinfo{author}{\bibfnamefont{F.}~\bibnamefont{Siebel}},
  \bibinfo{author}{\bibfnamefont{J.~A.} \bibnamefont{Font}},
  \bibinfo{author}{\bibfnamefont{E.}~\bibnamefont{Muller}}, \bibnamefont{and}
  \bibinfo{author}{\bibfnamefont{P.}~\bibnamefont{Papadopoulos}},
  \bibinfo{journal}{Phys. Rev. D} \textbf{\bibinfo{volume}{65}},
  \bibinfo{pages}{064038} (\bibinfo{year}{2002}).

\bibitem[{\citenamefont{Finn and Thorne}(2000)}]{FinnThorne}
\bibinfo{author}{\bibfnamefont{L.~S.} \bibnamefont{Finn}} \bibnamefont{and}
  \bibinfo{author}{\bibfnamefont{K.~S.} \bibnamefont{Thorne}},
  \bibinfo{journal}{Phys. Rev. D} \textbf{\bibinfo{volume}{62}},
  \bibinfo{pages}{124021} (\bibinfo{year}{2000}).

\bibitem[{\citenamefont{Bishop et~al.}(2003)\citenamefont{Bishop, G\'omez,
  Husa, Lehner, and Winicour}}]{particle}
\bibinfo{author}{\bibfnamefont{N.~T.} \bibnamefont{Bishop}},
  \bibinfo{author}{\bibfnamefont{R.}~\bibnamefont{G\'omez}},
  \bibinfo{author}{\bibfnamefont{S.}~\bibnamefont{Husa}},
  \bibinfo{author}{\bibfnamefont{L.}~\bibnamefont{Lehner}}, \bibnamefont{and}
  \bibinfo{author}{\bibfnamefont{J.}~\bibnamefont{Winicour}}
  (\bibinfo{year}{2003}), \bibinfo{note}{preprint gr-qc/0301060}.

\bibitem[{\citenamefont{G\'omez et~al.}(2002)\citenamefont{G\'omez, Husa,
  Lehner, and Winicour}}]{white}
\bibinfo{author}{\bibfnamefont{R.}~\bibnamefont{G\'omez}},
  \bibinfo{author}{\bibfnamefont{S.}~\bibnamefont{Husa}},
  \bibinfo{author}{\bibfnamefont{L.}~\bibnamefont{Lehner}}, \bibnamefont{and}
  \bibinfo{author}{\bibfnamefont{J.}~\bibnamefont{Winicour}},
  \bibinfo{journal}{Phys. Rev. D} \textbf{\bibinfo{volume}{66}},
  \bibinfo{pages}{064019} (\bibinfo{year}{2002}).

\bibitem[{\citenamefont{Campanelli et~al.}(2001)\citenamefont{Campanelli,
  G\'omez, Husa, Winicour, and Zlochower}}]{close}
\bibinfo{author}{\bibfnamefont{M.}~\bibnamefont{Campanelli}},
  \bibinfo{author}{\bibfnamefont{R.}~\bibnamefont{G\'omez}},
  \bibinfo{author}{\bibfnamefont{S.}~\bibnamefont{Husa}},
  \bibinfo{author}{\bibfnamefont{J.}~\bibnamefont{Winicour}}, \bibnamefont{and}
  \bibinfo{author}{\bibfnamefont{Y.}~\bibnamefont{Zlochower}},
  \bibinfo{journal}{Phys. Rev. D} \textbf{\bibinfo{volume}{63}},
  \bibinfo{pages}{124013} (\bibinfo{year}{2001}).

\bibitem[{\citenamefont{Husa et~al.}(2002)\citenamefont{Husa, Zlochower,
  G\'omez, and Winicour}}]{close2}
\bibinfo{author}{\bibfnamefont{S.}~\bibnamefont{Husa}},
  \bibinfo{author}{\bibfnamefont{Y.}~\bibnamefont{Zlochower}},
  \bibinfo{author}{\bibfnamefont{R.}~\bibnamefont{G\'omez}}, \bibnamefont{and}
  \bibinfo{author}{\bibfnamefont{J.}~\bibnamefont{Winicour}},
  \bibinfo{journal}{Phys. Rev. D} \textbf{\bibinfo{volume}{65}},
  \bibinfo{pages}{084034} (\bibinfo{year}{2002}).

\bibitem[{\citenamefont{Papadopoulos}(2002)}]{philipharm}
\bibinfo{author}{\bibfnamefont{P.}~\bibnamefont{Papadopoulos}},
  \bibinfo{journal}{Phys. Rev. D} \textbf{\bibinfo{volume}{65}},
  \bibinfo{pages}{084016} (\bibinfo{year}{2002}).

\bibitem[{\citenamefont{Zlochower et~al.}(2003)\citenamefont{Zlochower,
  G\'omez, Husa, Lehner, and Winicour}}]{mode}
\bibinfo{author}{\bibfnamefont{Y.}~\bibnamefont{Zlochower}},
  \bibinfo{author}{\bibfnamefont{R.}~\bibnamefont{G\'omez}},
  \bibinfo{author}{\bibfnamefont{S.}~\bibnamefont{Husa}},
  \bibinfo{author}{\bibfnamefont{L.}~\bibnamefont{Lehner}}, \bibnamefont{and}
  \bibinfo{author}{\bibfnamefont{J.}~\bibnamefont{Winicour}}
  (\bibinfo{year}{2003}), \bibinfo{note}{preprint gr-qc/0306098}.

\bibitem[{\citenamefont{G\'omez}(2001)}]{alterprd}
\bibinfo{author}{\bibfnamefont{R.}~\bibnamefont{G\'omez}},
  \bibinfo{journal}{Phys. Rev. D} \textbf{\bibinfo{volume}{64}},
  \bibinfo{pages}{024007} (\bibinfo{year}{2001}).

\bibitem[{\citenamefont{Friedrich}(1982)}]{friedrich}
\bibinfo{author}{\bibfnamefont{H.}~\bibnamefont{Friedrich}},
  \bibinfo{journal}{Proc. Roy. Soc. Lond.} \textbf{\bibinfo{volume}{A381}},
  \bibinfo{pages}{361} (\bibinfo{year}{1982}).

\bibitem[{\citenamefont{Stewart}(1990)}]{stewart}
\bibinfo{author}{\bibfnamefont{J.}~\bibnamefont{Stewart}},
  \emph{\bibinfo{title}{Advanced General Relativity}}
  (\bibinfo{publisher}{Cambridge University Press},
  \bibinfo{address}{Cambridge}, \bibinfo{year}{1990}).

\bibitem[{\citenamefont{Frittelli and Lehner}(1999)}]{sfll}
\bibinfo{author}{\bibfnamefont{S.}~\bibnamefont{Frittelli}} \bibnamefont{and}
  \bibinfo{author}{\bibfnamefont{L.}~\bibnamefont{Lehner}},
  \bibinfo{journal}{Phys. Rev. D} \textbf{\bibinfo{volume}{59}},
  \bibinfo{pages}{084012} (\bibinfo{year}{1999}).

\bibitem[{\citenamefont{Duff}(1958)}]{duff}
\bibinfo{author}{\bibfnamefont{G.~F.~D.} \bibnamefont{Duff}},
  \bibinfo{journal}{Can. J. Math.} \textbf{\bibinfo{volume}{10}},
  \bibinfo{pages}{127} (\bibinfo{year}{1958}).

\bibitem[{\citenamefont{LeVeque}(1990)}]{leveque}
\bibinfo{author}{\bibfnamefont{R.~J.} \bibnamefont{LeVeque}},
  \emph{\bibinfo{title}{Numerical Methods for Conservation Laws}}
  (\bibinfo{publisher}{Birkhauser-Verlag}, \bibinfo{address}{Basel},
  \bibinfo{year}{1990}).

\bibitem[{\citenamefont{LeVeque}(1997)}]{wave}
\bibinfo{author}{\bibfnamefont{R.~J.} \bibnamefont{LeVeque}},
  \bibinfo{journal}{J. Comput. Physics} \textbf{\bibinfo{volume}{131}},
  \bibinfo{pages}{327} (\bibinfo{year}{1997}).

\bibitem[{\citenamefont{Font}(2000)}]{fontreview}
\bibinfo{author}{\bibfnamefont{J.~A.} \bibnamefont{Font}},
  \bibinfo{journal}{Living Rev. Rel.} \textbf{\bibinfo{volume}{3}},
  \bibinfo{pages}{2} (\bibinfo{year}{2000}), \eprint{gr-qc/0003101}.

\bibitem[{\citenamefont{G\'omez et~al.}(1994)\citenamefont{G\'omez,
  Papadopoulos, and Winicour}}]{axisymmetric}
\bibinfo{author}{\bibfnamefont{R.}~\bibnamefont{G\'omez}},
  \bibinfo{author}{\bibfnamefont{P.}~\bibnamefont{Papadopoulos}},
  \bibnamefont{and} \bibinfo{author}{\bibfnamefont{J.}~\bibnamefont{Winicour}},
  \bibinfo{journal}{J. Math. Phys.} \textbf{\bibinfo{volume}{35}},
  \bibinfo{pages}{4184} (\bibinfo{year}{1994}).

\bibitem[{\citenamefont{Pons et~al.}(1998)\citenamefont{Pons, Font,
  Ib{\'a}{\~n}ez, Marti, and Miralles}}]{pons}
\bibinfo{author}{\bibfnamefont{J.}~\bibnamefont{Pons}},
  \bibinfo{author}{\bibfnamefont{J.}~\bibnamefont{Font}},
  \bibinfo{author}{\bibfnamefont{J.}~\bibnamefont{Ib{\'a}{\~n}ez}},
  \bibinfo{author}{\bibfnamefont{J.}~\bibnamefont{Marti}}, \bibnamefont{and}
  \bibinfo{author}{\bibfnamefont{J.}~\bibnamefont{Miralles}},
  \bibinfo{journal}{Astron. Astrophys.} \textbf{\bibinfo{volume}{339}},
  \bibinfo{pages}{638} (\bibinfo{year}{1998}).

\bibitem[{\citenamefont{Marti and M{\"u}ller}(1996)}]{marti}
\bibinfo{author}{\bibfnamefont{J.}~\bibnamefont{Marti}} \bibnamefont{and}
  \bibinfo{author}{\bibfnamefont{E.}~\bibnamefont{M{\"u}ller}},
  \bibinfo{journal}{J. Comput. Phys.} \textbf{\bibinfo{volume}{123}},
  \bibinfo{pages}{1} (\bibinfo{year}{1996}).

\bibitem[{\citenamefont{Siebel et~al.}(2003)\citenamefont{Siebel, Font, Muller,
  and Papadopoulos}}]{siebel}
\bibinfo{author}{\bibfnamefont{F.}~\bibnamefont{Siebel}},
  \bibinfo{author}{\bibfnamefont{J.~A.} \bibnamefont{Font}},
  \bibinfo{author}{\bibfnamefont{E.}~\bibnamefont{Muller}}, \bibnamefont{and}
  \bibinfo{author}{\bibfnamefont{P.}~\bibnamefont{Papadopoulos}},
  \bibinfo{journal}{Phys. Rev. D} \textbf{\bibinfo{volume}{67}},
  \bibinfo{pages}{124018} (\bibinfo{year}{2003}).

\bibitem[{\citenamefont{Pretorius and Lehner}(2003)}]{pretorius}
\bibinfo{author}{\bibfnamefont{F.}~\bibnamefont{Pretorius}} \bibnamefont{and}
  \bibinfo{author}{\bibfnamefont{L.}~\bibnamefont{Lehner}}
  (\bibinfo{year}{2003}), \bibinfo{note}{preprint gr-qc/0302003}.

\bibitem[{\citenamefont{Rossmanith et~al.}(2003)\citenamefont{Rossmanith, Bale,
  and LeVeque}}]{curvewave}
\bibinfo{author}{\bibfnamefont{J.~A.} \bibnamefont{Rossmanith}},
  \bibinfo{author}{\bibfnamefont{D.~S.} \bibnamefont{Bale}}, \bibnamefont{and}
  \bibinfo{author}{\bibfnamefont{R.~J.} \bibnamefont{LeVeque}}
  (\bibinfo{year}{2003}), \bibinfo{note}{preprint}.

\bibitem[{\citenamefont{G\'omez et~al.}(1997)\citenamefont{G\'omez, Lehner,
  Papadopoulos, and Winicour}}]{eth}
\bibinfo{author}{\bibfnamefont{R.}~\bibnamefont{G\'omez}},
  \bibinfo{author}{\bibfnamefont{L.}~\bibnamefont{Lehner}},
  \bibinfo{author}{\bibfnamefont{P.}~\bibnamefont{Papadopoulos}},
  \bibnamefont{and} \bibinfo{author}{\bibfnamefont{J.}~\bibnamefont{Winicour}},
  \bibinfo{journal}{Class. Quantum Grav.} \textbf{\bibinfo{volume}{14}},
  \bibinfo{pages}{977} (\bibinfo{year}{1997}).

\bibitem[{\citenamefont{Alekseenko and Arnold}(2002)}]{alekseenko}
\bibinfo{author}{\bibfnamefont{A.~M.} \bibnamefont{Alekseenko}}
  \bibnamefont{and} \bibinfo{author}{\bibfnamefont{D.~N.} \bibnamefont{Arnold}}
  (\bibinfo{year}{2002}), \bibinfo{note}{preprint gr-qc/0210071}.

\bibitem[{\citenamefont{York}(1979)}]{yorksources}
\bibinfo{author}{\bibfnamefont{J.}~\bibnamefont{York}}, in
  \emph{\bibinfo{booktitle}{Sources of Gravitational Radiation}}, edited by
  \bibinfo{editor}{\bibfnamefont{L.}~\bibnamefont{Smarr}}
  (\bibinfo{publisher}{Cambridge University Press},
  \bibinfo{address}{Cambridge}, \bibinfo{year}{1979}), p.~\bibinfo{pages}{83}.

\end{thebibliography}
\end{document}